\documentclass[10pt, prd, twocolumn]{revtex4}

\usepackage{amsmath}
\usepackage{amssymb}
\usepackage{graphicx}
\usepackage{bm} 
\usepackage{xcolor}

\begin{document}

\title{Thermodynamics of five-dimensional
Schwarzschild black holes
in the canonical ensemble}

\author{Rui Andr\'{e}}
\email{rui.andre@tecnico.ulisboa.pt}
\affiliation{Centro de Astrof\'{\i}sica e Gravita\c c\~ao  - CENTRA,
Departamento de F\'{\i}sica, Instituto Superior T\'ecnico - IST,
Universidade de Lisboa - UL, Av. Rovisco Pais 1, 1049-001 Lisboa, Portugal}
\author{Jos\'{e} P. S. Lemos}
\email{joselemos@tecnico.ulisboa.pt}
\affiliation{Centro de Astrof\'{\i}sica e Gravita\c c\~ao  - CENTRA,
Departamento de F\'{\i}sica, Instituto Superior T\'ecnico - IST,
Universidade de Lisboa - UL, Av. Rovisco Pais 1, 1049-001 Lisboa, Portugal}

\begin{abstract}

We study the thermodynamics of a five-dimensional Schwarzschild black
hole, also known as a five-dimensional Schwarzschild-Tangherlini black
hole, in the canonical ensemble using York's formalism.  Inside a
cavity of fixed size $r$ and fixed temperature $T$, we find that there
is a threshold at $\pi r T = 1$ above which a black hole can be in
thermal equilibrium with the cavity's boundary.  Moreover, this
thermal equilibrium can only be achieved for two specific types of
black holes.  One is a small, compared with the cavity size $r$, black
hole of horizon radius $r_{+1}$, while the other is a large, of the
order of the cavity size $r$, black hole of horizon radius $r_{+2}$.
In five dimensions, the radii $r_{+1}$ and $r_{+2}$ have an exact
expression.  Through the path integral formalism, which directly
yields the partition function of the system, one obtains the action
and thus the free energy for the black hole in the canonical ensemble.
The procedure leads naturally to the thermal energy and entropy of the
canonical system, the latter turning out to be given by the
Bekenstein-Hawking area law $S = \frac{A_{+}}{4}$, where the black
hole's surface area in five dimensions is $A_+ = 2\pi^2 r_+^3$ and
$r_+$ stands for both $r_{+1}$ and $r_{+2}$.
We also calculate the heat capacity and find that it is positive when
the heat bath is placed at a radius $r$ that is equal or less than the
photonic orbit, implying in this case thermodynamic stability, and
instability otherwise.  This means that the small black hole $r_{+1}$
is unstable and the large one $r_{+2}$ is stable.  A generalized free
energy is used to compare the possible thermodynamic phase transitions
relative to classical hot flat space which has zero free energy, and
we show that it is feasible in certain instances that classical hot
flat space transits through $r_{+1}$ to settle at the stable $r_{+2}$,
with the free energy of the unstable smaller black hole $r_{+1}$
acting as the potential barrier between the two states. It is also
shown that, remarkably, the free energy of the larger $r_{+2}$ black
hole is zero when the cavity radius is equal to the Buchdahl radius.
The relation to the instabilities that arise due to perturbations in
the path integral in the instanton solution is mentioned.  Hot flat
space is made of gravitons and it should be treated quantum
mechanically rather than classically.  Quantum hot flat space has
negative free energy and we find the conditions for which the large
black hole phase, quantum hot flat space phase, or both are the ground
state of the canonical ensemble. The corresponding phase diagram is
displayed in a $r\times T$ plot showing clearly the three possible
phases.  Using the density of states $\nu$ at a given energy $E$ we
also find that the entropy of the large black hole $r_{+2}$ is $S =
\frac{A_{+2}}{4}$. In addition, we make the connection between the
five-dimensional thermodynamics and York's four-dimensional results.

\end{abstract}


\maketitle



\newpage

\section{Introduction}

York substantiated the study of black hole thermodynamics by
studying a four-dimensional Schwarzschild  black hole in the
canonical ensemble and deriving the corresponding Euclidean path
integral \cite{york1}.  In this formalism the black hole is put in a
heat bath, i.e., inside a spherical cavity that
has a radius $r$ at temperature $T$.

This stemmed from the work of Hawking and collaborators
\cite{hawking2,hawking3,hawking4} that used path-integral techniques
to study gravitational systems containing black holes and to show that,
by imposing precise boundary data on the corresponding Euclidean
spacetime instanton, the black hole temperature is fixed through the
correct period in imaginary time.
With these techniques it was found that, in Planck
units, the entropy $S$ is indeed $S=\frac14A_+$, where $A_+$ is the
black hole
horizon area, which comes from the contribution of the classical
first-order Euclidean Einstein action of the black hole geometry to
the canonical partition function.  In ordinary quantum field theory
the value of the classical action is absorbed into the normalization
of the functional integral and so is physically irrelevant, but in
gravitational theories and, in particular, in black hole spacetimes the
classical action contributes to the entropy.  It was stressed in
\cite{hawking4} that in the path-integral approach the Euler
characteristic $\chi$ of the four-dimensional Euclidean black hole,
which gives its topology, is the feature that gives a
nonzero entropy.

One can then perturb the instanton solution of
the Euclidean classical black hole through thermal quantum
fluctuations and find that, while
for a large cavity radius the solution is
unstable \cite{gpy}, for a small cavity radius the
solution is stable \cite{allen}.  This came full circle with the
thermodynamic work of York \cite{york1}, where
in the canonical ensemble
a small black hole in a given
cavity is unstable, whereas
a large black hole 
is stable. With hindsight one finds 
that Hawking's path-integral approach generated the correct black hole
entropy, but the formalism available \cite{hawking2,hawking3,hawking4}
only applied for the small black hole, which is unstable and thus
defies a proper thermodynamic treatment.

York's method has been studied for different black holes, such as
charged black holes in the grand canonical
ensemble~\cite{yorkmartinez,york2}, black holes with matter in
arbitrary configurations \cite{zaslavskii1}, and black holes in
anti-de Sitter spacetimes~\cite{pecalemos}.

Another direction of study is to extend the York and path-integral
perturbation approaches to higher dimensions \cite{GregRoss}, see also
\cite{reallbranes,lu}. Correspondingly, the path-integral approach was
also used to extend to higher dimensions the idea that the black hole
entropy stems from the Euler characteristic $\chi$ of a Euclidean
black hole.  Indeed, in \cite{btz1994} it was shown that the entropy
of a black hole in any dimension stems solely from the Euler
characteristic of the two-dimensional plane~$(r,t)$, spanned by the
radial coordinate $r$ and Euclidean time $t$.

In this work we perform a full study of the thermodynamics of the
five-dimensional Schwarzschild black hole, also called a
Schwarzschild-Tangherlini black
hole, in a canonical ensemble using York's
ideas developed for the four-dimensional Schwarzschild black hole
\cite{york1}.  The important higher-dimensional analysis done in
\cite{GregRoss} was more concerned with the stable and unstable modes
of the black holes in the path-integral approach.  Among several
results in our thermodynamic analysis, we find the ground state of the
canonical ensemble, explicitly showing for which conditions the ground
state is provided by the large black hole or by quantum hot flat
space. We make use of the Schwarzschild-Tangherlini metric in five
dimensions \cite{tangherlini}, we mention the Buchdahl limiting radius
in four \cite{buchdahl} and higher dimensions \cite{wright}, and cite
the photonic orbit radius of the five-dimensional
Schwarzschild-Tangherlini metric \cite{monteiro}.  For the quantum hot
flat space properties in five dimensions that we need, see
\cite{landsberg}.

The paper is organized as follows.  In Sec.~\ref{secCavity} we
construct a cavity with an interior black hole, and by fixing the
cavity's size and temperature we find the possible black hole
solutions that are in thermal equilibrium with the surrounding heat
bath.  We also compute the Euclidean Einstein-Hilbert action for a
five-dimensional Schwarzschild black hole as a function of the
cavity's temperature and size.  In Sec.~\ref{secThermo} we derive all
relevant thermodynamic quantities of the black hole, write the first
law of thermodynamics, and find the thermodynamic energy, 
pressure, and  entropy of the system.  In Sec.~\ref{secStability}
we compute the heat capacity in order to find which of the black hole
solutions in thermal equilibrium are actually stable. In
Sec.~\ref{secFreeEnergyFunc} we study a generalized Helmholtz free
energy function, which is the thermodynamic potential of the canonical
ensemble proportional to the action, and elaborate on its role in
black hole nucleation from classical hot flat space, studying the
corresponding instabilities.  In Sec.~\ref{sa} we comment on the
action functional to second order and its relation to thermodynamics
and thermal stability.  In Sec.~\ref{secGround} we treat hot flat
space in quantum terms as a bath of gravitons, and identify in which
situations the black hole solution phase, the quantum hot flat space
phase, or the coexistence of both phases is the ground state of the
canonical ensemble, i.e., which of the phases has the lowest free
energy.  In Sec.~\ref{secDensity} we find
the density of states for the stable black hole solution
using the partition function and manage to
compute the entropy of the stable black hole.  In
the Appendix we comment on York's four-dimensional results
and compare them with the five-dimensional results.

\section{The canonical ensemble for a cavity with a black hole inside:
Temperature,
the Euclidean Einstein action,
and the action functional or partition function
for a five-dimensional Schwarzschild black hole}
\label{secCavity}

\subsection{The cavity in five dimensions and
the canonical temperature}

The temperature at infinity of a black hole is defined by 
Hawking's prescription for quantum fields in curved spacetimes.
We use units in which the
gravitational constant, the velocity of light,
the Planck constant, and the Boltzmann
constant are set to unity,
$G=1$, $c=1$,  $\hbar=1$, and $k_B=1$, respectively.
The Schwarzschild
Euclidean metric in five dimensions is given
in the usual spherical coordinates
by
the line element \cite{tangherlini}
\begin{equation}\label{metric}
ds^2 = \left(1 - \frac{r_+^2}{r^2} \right) dt^2 + 
\left(1 - \frac{r_+^2}{r^2} \right)^{-1}
dr^2 +r^2 d\Omega^2_{3},
\end{equation} 
where $t$ is the Euclidean imaginary time,
$r$ is the radial coordinate,
the subscript 3 in $\Omega_{3}$
indicates the three
angular coordinates $\theta_1$, $\theta_2$, $\theta_3$, with
$d\Omega^2_{3}=d\theta_1^2+\sin^2\theta_1
\left(d\theta_2^2+\sin^2\theta_2d\theta_3^2\right)$,
and $r_+$ is the
gravitational radius related
in five dimensions to the spacetime's ADM 
mass $m$ through $r_+^2=\frac{8}{3\pi}\,m$. In general, we
always use $r_+$, not $m$.
The metric describes
the space outside of the event horizon $r_+$
and is devoid of singularities,
provided that the a priori
conical singularity at the horizon is removed by
attributing the correct period to the Euclidean time.
Close to the horizon, by redefining $r=r_+ + \varepsilon$ for small
$\varepsilon$ and introducing $\rho = \sqrt{2 r_+ \varepsilon}$,
the metric in the reduced $(t,\rho)$ Euclidean space
is
$
ds^2 = \rho^2 \left( \frac{dt}{r_+} \right)^2 + d\rho^2 
$.
This equation clearly has a Euclidean polar coordinate
form,
where the angular coordinate is $t/r_+$, and so 
to avoid conical singularities $t$ must have period
$2\pi r_+$. Thus, the Euclidean two-space has  topology
$R^2$ and the full Euclidean space has  topology
$R^2 \times S^3$.
We denote the Euclidean time period by $\beta_\infty$, so
$\beta_\infty = 2\pi r_+$.
From Hawking's prescription,
this is the inverse  Hawking temperature $T_\infty=
\frac{1}{\beta_\infty}=\frac{1}{2\pi r_+}$.

In statistical physics, the canonical ensemble
is defined as the ensemble for which the temperature $T$
is kept fixed.
Since our gravitational
system is spherically symmetric it is natural
to define the canonical ensemble as 
one for which the temperature $T$ is fixed at some
boundary radius $r_{\rm B}$,
which henceforth we abbreviate to $r$, and which should
not be confounded with the  usual radial
coordinate $r$.
This boundary at radius $r$
has topology $S^1 \times S^3$, with 
$S^3$ having the usual 
three-surface area $2\pi^2 r^3$.
From Eq.~(\ref{metric}), $S^1$
has proper length $\beta$ given by
$\beta = \int_0^{\beta_\infty} \sqrt{1 - \frac{r_+^2}{r^2}} dt
=\beta_\infty\sqrt{1 - \frac{r_+^2}{r^2}}$, which upon
using $\beta_\infty = 2\pi r_+$ yields the length
\begin{equation}\label{beta}
\beta =
2\pi r_+ \sqrt{1 - \frac{r_+^2}{r^2}}.
\end{equation}
Imposing an isothermal boundary condition at $r$ is
the same as imposing
a fixed proper length $\beta$.
To see this, we define the local temperature $T$ as 
\begin{equation}\label{betat}
T = \frac{1}{\beta}\,,
\end{equation}
so that $\beta$ is the inverse temperature.
Then, from $\beta =\beta_\infty\sqrt{1 - \frac{r_+^2}{r^2}}$
one finds for $T$ that 
$T =\frac{T_\infty}
{\sqrt{1 - \frac{r_+^2}{r^2}}}$, 
which is the Tolman temperature formula
for the system formed by the black hole and cavity at $r$.
This states that the temperature at the
cavity is the temperature of the black hole
at infinity, i.e., the Hawking temperature, blueshifted
to the cavity's position $r$, or conversely,
the temperature  of the black hole
at infinity, i.e., the Hawking temperature, is the temperature
$T$ at the cavity redshifted to infinity.
Using 
$T_\infty=\frac{1}{\beta_\infty}=\frac{1}{2\pi r_+}$,
we can  write
the canonical temperature as
\begin{equation}\label{tinverbeta}
T =\frac{1}
{2\pi r_+ \sqrt{1 - \frac{r_+^2}{r^2}}}\,.
\end{equation}

Given a temperature $T$ and a cavity radius $r$,
we want to know
which values of black hole horizon radii $r_+$ are allowed,
i.e.,
we need to know $r_+=r_+(\beta,r)$ or
$r_+=r_+(T,r)$. Note that we exchange $T$ and $\beta$ at will.
For that we solve the
Tolman condition for temperature variation in curved space, which is
given by Eq.~(\ref{tinverbeta}). Equation~(\ref{tinverbeta}) can be
rewritten as
\begin{equation}
\label{bhequation}
\left( 2\pi T \right)^2 \, r_+^4 -
\left( 2\pi T \right)^2 r^2 \, r_+^2 + r^2 = 0\,.
\end{equation}
Equation~(\ref{bhequation}) is a quadratic
equation for $r_+^2$
and can be solved exactly.
In York's $d=4$ case, it is a cubic equation.
Equation~(\ref{bhequation}) only has real solutions for
\begin{equation}\label{real}
\pi rT \geq 1\,.
\end{equation}
If we fix the radius $r$ of the cavity, then a black hole can form only if
the temperature $T$ is high enough; otherwise, one has empty hot flat
space, which if one performs a quantum treatment means hot gravitons
wandering around in flat space.  On the other hand, if we fix $T$ at
the cavity, then only if the radius is high enough can a black hole
develop; otherwise, for low $r$ one has only empty hot flat space.
Equation~(\ref{real}) is a necessary condition for having black holes,
but not sufficient.

The two possible 
black hole solutions $r_{+1}$ and $r_{+2}$
of Eq.~(\ref{bhequation}) are given by 
\begin{align}
r_{+1} & = r \left(
\frac{1 - \sqrt{1 - (\pi r T)^{-2}}}{2} \right)^{1/2}, \label{bh1}\\ 
r_{+2} & = r
\left( \frac{1 + \sqrt{1 - (\pi r T)^{-2}}}{2} \right)^{1/2}
\label{bh2},
\end{align}
with $r_{+2} \geq r_{+1}$.
For a connection to four dimensions, see the Appendix.
\begin{figure}[h]
\centering
\includegraphics[width=0.48\textwidth]{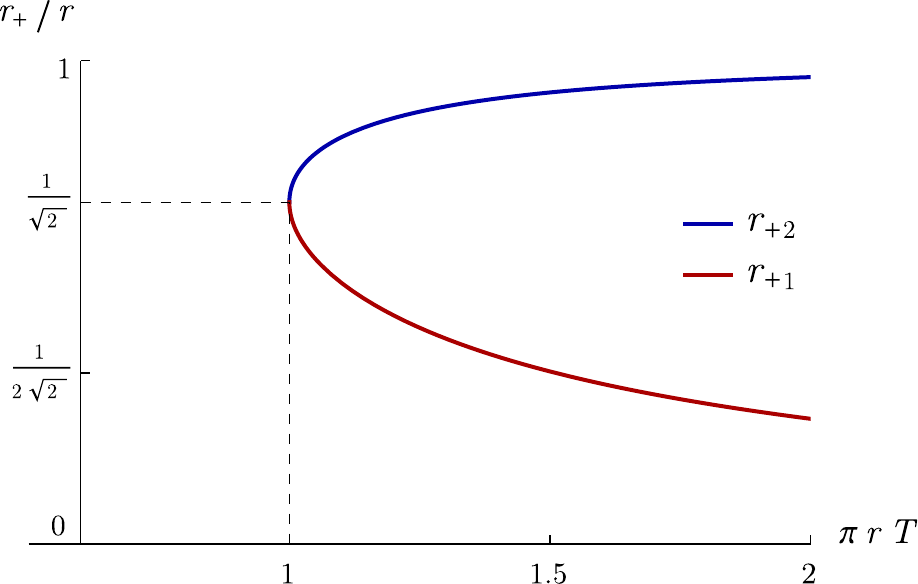}
\caption{The two black hole solutions $r_{+1}$ and $r_{+2}$ are
represented for different parameter values of $\pi r T \geq 1$.  The
two coincide for $\pi r T= 1$ at $\frac{r_+}r = \frac{1}{\sqrt{2}}$,
represented by a point at the intersection of the dashed lines.  The
radius $r$ given by $\frac{r_+}r = \frac{1}{\sqrt{2}}$ is the radius
of the photon sphere $r_{\rm ph} = \sqrt2 \, r_+$.  It turns out that
the $r_{+1}$ solutions are thermodynamically unstable and the $r_{+2}$
solutions are thermodynamically stable.  }
\label{BHs}
\end{figure}
Equations~(\ref{bh1}) and (\ref{bh2}) give
$r_+=r_+(r,\beta)$ or $r_+=r_+(r,T)$
for each black hole,
i.e., they give the horizon radius
of each black hole as a function of the
cavity radius $r$ and either 
the inverse temperature $\beta$
or  temperature $T$.
In Fig.~\ref{BHs}
we
plot 
$r_{+1}$ and $r_{+2}$ as
functions of $T$, given in 
Eqs.~(\ref{bh1}) and (\ref{bh2}).
Note from Eqs.~(\ref{bh1}) and (\ref{bh2}),
and also from Fig.~\ref{BHs},
that for low temperatures, i.e., for
$ \pi r T = 1$ or near this value,
we have $r_{+1}=r_{+2} = \frac{r}{\sqrt2} $. 
For
high temperatures,
as $\pi r T \rightarrow \infty$,
from Eqs.~(\ref{bh1}) and~(\ref{bh2})
one
has
$
r_{+1} = r \left(
\frac{1}{2\pi r T} + \mathcal{O}
\left(\frac{1}{(\pi r T)^{3}}\right)
\right)
$,
and
$
r_{+2} = r \left( 1 -
\frac{1}{8 \left(\pi r T \right)^2} +
 \mathcal{O}
\left(\frac{1}{(\pi r T)^{4}}\right)
\right)
$,
so that in the limit 
the smaller black hole 
is given by $r_{+1}=0$, while
the larger one approaches the cavity's 
boundary $r_{+2}= r$. 
The radii 
$r_{+1}$ and $r_{+2}$ in 
Eqs.~(\ref{bh1}) and (\ref{bh2})
are the radii that we need to put
into the action 
to get the final action $I=I(r,\beta)$,
to which we now turn.

\subsection{The Euclidean-Einstein action
and the action functional or the partition function
for a five-dimensional Schwarzschild black hole}
\label{secAction}

In the Hartle-Hawking path-integral approach to quantum gravity,
the partition function $Z$ for the gravitational field is given by the
integral of the Euclidean Einstein action $I$ over the space of
metrics
$\rm g$, i.e., $Z= \int d[\rm g] \, {\rm e}^{-I[\rm g]}$.
Assuming that the leading term is that of the first-order classical
Euclidean action of a black hole, $I_{\rm BH}$, one has
\begin{equation}\label{zeroloop}
Z= {\rm e}^{-I_{\rm BH}}.
\end{equation}

The Euclidean Einstein action for a five-dimensional spacetime on a
compact region $\mathcal M$ is given by
\begin{equation}\label{a1}
I = - \frac1{16\pi} \int_\mathcal{M}d^5 x \sqrt{|g|}R - \frac1{8\pi} 
\int_{\partial M} d^{4}x \sqrt{|h|} [K]\,,
\end{equation}
where $|g|$ is the determinant
of the metric $g_{ab}$,
with $a,b$ running through one time
and four spatial coordinates,
$R$ is the Ricci scalar,
$|h|$ is the determinant of the induced metric on the four-dimensional
boundary $\partial \mathcal M$, and $[K]$ is
the difference between the 
trace of the extrinsic curvature $K$ of the hypersurface in the metric 
$g$ and the extrinsic curvature of the hypersurface in a
reference spacetime metric, which here is 
the Minkowski spacetime metric,  $K_{\rm flat}$.
This subtraction is responsible for setting the action to zero in
the reference flat spacetime. The first part of the action~(\ref{a1})
is the
bulk term, and the second part is
the Gibbons-Hawking-York boundary term.

Now, for the metric Eq.~(\ref{metric})
the Ricci scalar vanishes everywhere for the Schwarzschild
metric, $R=0$, and so
the first term in the action of Eq.~(\ref{a1}) vanishes. The action
then reduces 
to the Gibbons-Hawking-York boundary term
$
I = - \frac{1}{8\pi} \int_{\partial \mathcal M} d^4 x \sqrt{|h|} [K]
$.
The induced line element
on $r={\rm constant}$ from Eq.~(\ref{metric}) is 
$ds^2\vert_r= \left(1 - \frac{r_+^2}{r^2} \right)
dt^2 +r^2 d\Omega^2_{3}$,
the determinant $h$ of the induced
metric is 
$h=\left(1- \frac{r_+^2}{r^2}\right)
r^6\sin^4\theta_1\sin^2\theta_2$, and 
the trace of the boundary's extrinsic curvature is
$
K = \frac{3\sqrt{1 - \frac{r_+^2}{r^2}}}{r} + 
\frac{r_+^2}{r^3 \sqrt{1 - \frac{r_+^2}{r^2}}}
$,
while its flat-space counterpart can be
found directly by putting $r_+=0$,
$
K_{\rm flat} = \frac3r
$.
Since at the boundary $r$ is constant,
one only has to integrate over the angles, which gives
$2\pi^2$, and integrate over $t$
from 0 to $\beta_\infty$.
The action $I$,
Eq.~(\ref{a1}),
takes the final form
\begin{equation}\label{finalaction}
I =  \pi^2 r_+^3 - \frac{3}{2}\pi^2r_+ r^2 +
\frac{3\pi^2}{2}
r_+ \sqrt{1 - \frac{r_+^2}{r^2}}\,r^2,
\end{equation}
so that, in this form, $I=I(r,r_+)$.

We are interested in writing
the action~(\ref{finalaction})
as a function of $r$ and $\beta$
only, $I=I(r,\beta)$.
This can indeed be done
since $r_+= r_+(r,\beta)$,
and Eq.~(\ref{finalaction}) is then formally 
\begin{equation}\label{action5d}
I(r,\beta) =  \pi^2 r_+^3(r,\beta) -
\frac{3}{2}\pi^2r_+(r,\beta) r^2 +
\frac{3\pi}{4}\beta r^2\,,
\end{equation}
where $r_+$ stands for
$r_{+1}$ or $r_{+2}$ given in 
Eqs.~(\ref{bh1}) and (\ref{bh2}).
Thus, explicitly, 
\begin{align}
{I}(r,r_{+1}(r,\beta)) &= \frac{3 \pi^2 r^3}{4}
\Bigg[ (\pi rT)^{-1} -                           \nonumber \\
&
\frac{2\sqrt2}{3}\sqrt{1-\sqrt{1-(\pi rT)^{-2}}}\,\times \nonumber \\
&
\left( 1 + \frac{\sqrt{1-(\pi rT)^{-2}}}2 \right) \Bigg]\;,
0\leq \frac{r_{+1}}{r}\leq\frac{1}{\sqrt2}\,,
\label{bhactionr+1}
\end{align}
and
\begin{align}
{I}(r,r_{+2}(r,\beta)) &= \frac{3 \pi^2 r^3}{4} \Bigg[ (\pi rT)^{-1} 
-\nonumber \\
&-\frac{2\sqrt2}{3}\sqrt{1+\sqrt{1-(\pi rT)^{-2}}}\, \times\nonumber \\
&\left( 1- \frac{\sqrt{1-(\pi rT)^{-2}}}2 \right) \Bigg]\;,
\frac{1}{\sqrt2}\leq \frac{r_{+2}}{r}\leq1\,,
\label{bhactionr+2}
\end{align}
where $T=\frac1\beta$, see Eq.~(\ref{betat}).
To have black hole solutions, Eq.~(\ref{real}) must hold,
i.e., $\pi r T \geq1$.
For the smaller black hole $r_+ = r_{+1}$,
Eq.~(\ref{bh1}), the action~(\ref{bhactionr+1})
is always 
positive. 
For the larger black hole $r_+ = r_{+2}$,
Eq.~(\ref{bh2}), the action~(\ref{bhactionr+2}) is
positive  for 
$\pi r T < \frac{2}{\sqrt 3}$, i.e.,
$\pi r T < 1.155$ in round numbers, or
equivalently, for
$\frac{r_+}{r} < \frac{\sqrt3}{2}$, i.e.,
$\frac{r_+}{r} < 0.866$ in round numbers,
in brief, the action exists and is positive for
$1<\frac{r}{r_+} < \frac{2}{\sqrt3}$.
The action~(\ref{bhactionr+2}) is zero or negative otherwise,
namely, 
\begin{equation}
\label{negativeactionbuch}
\frac{r}{r_+} \geq \frac{2}{\sqrt3}\,. 
\end{equation}
Note that $\frac{2}{\sqrt3}$ is the Buchdahl bound in five dimensions.
In York's thermodynamic analysis in four dimensions \cite{york1} the
equivalent of Eq.~(\ref{negativeactionbuch}) is $\frac{r}{r_+} \geq
\frac{9}{8}$, and $\frac{9}{8}$ is precisely the Buchdahl bound in
four dimensions, a fact that went unnoticed up to now.  The Buchdahl
bound is a lower bound for the ratio of a star radius $r$ to its
gravitational radius $r_+$, that arises in order to maintain the star
solution free of singularities.  In four dimensions the bound is given
by $\frac{r}{r_+}\geq\frac{9}{8}$ \cite{buchdahl}, such that by
defining a four-dimensional Buchdahl limiting radius $r_{\rm
Buch}=\frac{9}{8}r_+$ one has $r\geq r_{\rm Buch}$, whereas in five
dimensions the bound is given by $\frac{r}{r_+}\geq\frac{2}{\sqrt3}$
\cite{wright}, such that by defining a five-dimensional Buchdahl
limiting radius $r_{\rm Buch}=\frac{2}{\sqrt3}r_+$ one has $r\geq
r_{\rm Buch}$. One sees that the Buchdahl limiting radius appears in
two different contexts, namely, in the original context of star
solutions and gravitational collapse in general relativity, and in the
new thermodynamic context of having a negative action and thus a
negative energy for a black hole in a canonical ensemble in general
relativity.

\section{Thermodynamics}
\label{secThermo}

To study thermodynamics in the canonical ensemble
of a black hole inside a cavity at
temperature $T$,
we connect the action
$I$ with the relevant thermodynamic potential,
namely, the free energy $F$. The connection is
given by\begin{equation}\label{actionoriginal}
I = \beta F\,.
\end{equation}
The action $I$ is taken as 
Eq.~(\ref{finalaction})
or in Eq.~(\ref{action5d}).
On the other hand, in thermodynamics
the free energy $F$ is related to
the thermodynamic energy $E$, the temperature $T$,
and the entropy $S$ by
\begin{equation}\label{energyf}
F = E - T\,S\,.
\end{equation}
The first law of thermodynamics
can be written then as
\begin{equation}\label{1stlaw}
dE = TdS-pdA\,,
\end{equation}
where
$A$
is the area of the cavity and $p$ is
the tangential pressure at the cavity
radius. 
The term $pdA$ is the correct term
to use in the first law, rather
than
a volume term, one of the reasons being that
the inside of
the black hole has no well-defined 
volume in this setting, see \cite{york1}.
So 
$A$ is indeed the independent variable and the tangential
pressure $p$ is conjugate to it.
The first law as written in Eq.~(\ref{1stlaw})
envisages the energy $E$ as a function
of the independent variables $S$ and $A$, i.e.,
$E=E(S,A)$, and $T$ and $p$ are state functions
that in general must be provided or somehow worked out,
$T=T(S,A)$ and $p=p(S,A)$, i.e., they are
the equations of state. It is also useful
to give the first law in a different set of variables.
Note from Eq.~(\ref{actionoriginal})
that $dI=\beta dF+Fd\beta$, and using
Eq.~(\ref{energyf}) for $dF$ one finds that the
first law in Eq.~(\ref{1stlaw})  can also be written as
\begin{equation}\label{1stlawI}
dI=Ed\beta-pdA\,,
\end{equation}
where here $I$ is 
taken  as a function
of the independent variables $\beta$ and $A$,
$I=I(A,\beta)$. In this form
one finds immediately that
\begin{align}
E &= \left( \frac{\partial I}{\partial \beta} \right)_A\,,
\label{energy1}\\
p & = -\frac1\beta \left( \frac{\partial I}{\partial A}
\right)_\beta\,,
\label{pressurei}\\
S &= \beta E - I\,.
\label{S1}
\end{align}
Note also that $I$  can be 
seen  as a function
of the independent variables $\beta$ and $r$, $I=I(r,\beta)$,
instead of $\beta$ and $A$, $I=I(A,\beta)$,
since a three-dimensional sphere has area $A$
given by 
\begin{equation}\label{area}
A= 2\pi^2 r^3\,.
\end{equation}
Let us now find $E$, $p$, and $S$ for the system.

To find $E$
we have to work out 
$\left( \frac{\partial I}{\partial \beta} \right)_A$,
and we proceed as follows. 
Envisaging $I=I(r,\beta)$,
one has
$dI = \left( \frac{\partial I}{\partial \beta} \right)_r d\beta +
\left( \frac{\partial I}{\partial r} \right)_\beta dr$.
Envisaging $I=I(r,r_+)$,
one has
$dI=
\left( \frac{\partial I}{\partial {r_+}} \right)_r dr_+ + 
\left( \frac{\partial I}{\partial r} \right)_{r_+} dr$.
Both $dI$s are the same, and so
at constant $r$ we can write
$\left( \frac{\partial I}{\partial \beta} \right)_r = 
\frac{(\partial I / \partial r_+)_r}{(\partial \beta /
\partial r_+)_r}$.
From Eq.~(\ref{finalaction}) we have
$\left( \frac{\partial I}{\partial r_+} \right)_r = 
\frac{3\pi^2r^2 \left( 1-\sqrt{1-r_+^2/r^2}
\right)}{2\sqrt{1-r_+^2/r^2}} 
\left(1 - \frac{2 r_+^2}{r^2} \right)$.
From Eq.~(\ref{beta}) we get
$
\left( \frac{\partial \beta}{\partial r_+} \right)_r =
\frac{2\pi}{\sqrt{1-r_+^2/r^2}} \left(1 -
\frac{2 r_+^2}{r^2} \right)$.
Then, from 
Eq.~(\ref{energy1}) we get the thermodynamic energy $E$,
\begin{equation}\label{E}
E = \frac{3\pi r^2}{4} \left(1- \sqrt{1-\frac{r_+^2}{r^2}} \right)\,.
\end{equation}
It is interesting to note
that the total thermodynamic energy $E$
inside the cavity is actually
larger than the spacetime's ADM mass.
Indeed, using $r_+^2=\frac{8}{3\pi}\,m$
together with Eq.~(\ref{E}), one finds
$
m = E - \frac{2 E^2}{3 \pi r^2}
$,
so that the  spacetime's total
energy is read as the
thermodynamic energy inside the cavity plus its binding 
energy.

To find the pressure from  Eq.~(\ref{pressurei}) we 
use $\left(\frac{\partial I}{\partial r} \right)_\beta =
\left(\frac{\partial I}{\partial r} \right)_{r_+} - 
\left(\frac{\partial I}{\partial \beta} \right)_r
\left(\frac{\partial \beta}{\partial r} \right)_{r_+}$,
together with Eqs.~(\ref{beta})~and~(\ref{action5d}).
This gives 
\begin{equation}\label{finalp}
p= \frac{1}{8 \pi r \sqrt{1-\frac{r_+^2}{r^2}} }
\left( 1-\sqrt{1-\frac{r_+^2}{r^2}} \right)^2\,.
\end{equation}

Now we can find the entropy $S$
of the system.
From Eq.~(\ref{S1}), together with
Eqs.~(\ref{action5d})
and (\ref{E}),
we find that the total entropy inside the cavity
becomes $S = \pi^2r_+^3/2$, i.e.,
\begin{equation}\label{finalentropy}
S = \frac{A_+}{4},
\end{equation}
which is the Bekenstein-Hawking
area law for the black hole entropy, 
with the five-dimensional surface area of
the black hole given by Eq.~(\ref{area})
putting $r=r_+$, i.e., 
$A_+ = 2\pi^2r_+^3$.
Here $r_+$ can be either $r_{+1}$ or
$r_{+2}$.
Knowing $S$, one can again find
$p$ using instead
the first law as written in Eq.~(\ref{1stlaw}).
Then, 
$p=-\left( \frac{\partial E}{\partial A} \right)_S$
and so
$
p =  \frac{1}{8 \pi r \sqrt{1-\frac{r_+^2}{r^2}} }
\left( 1-\sqrt{1-\frac{r_+^2}{r^2}} \right)^2
$, which coincides with Eq.~(\ref{finalp}), and
where Eq.~(\ref{E}) has been used.

One can also work out the 
Euler relation, the scaling laws, and the Gibbs-Duhem relation.
From the energy and entropy in Eqs.~(\ref{E}) and
(\ref{finalentropy}),
it is straightforward that
$
E=\frac{3\pi}{4}\left(\frac{A}{2\pi^2}\right)^{2/3} 
\left(1 - \sqrt{ 1 - \left( \frac{4S}{A} \right)^{2/3}} \right)
$.
Using Euler's theorem for homogeneous functions, one has that
$E$ is homogeneous of degree $\frac23$ in $A$ and $S$, i.e.,
$\frac{2}{3}E=
\left(\frac{\partial E}{\partial S}\right)S+
\left(\frac{\partial E}{\partial A}\right)A$, and the Euler
relation for the cavity is obtained as
\begin{equation}\label{euler equation}
\frac23 E =	TS - pA.
\end{equation}
The scaling laws for the canonical ensemble are thus 
$r \rightarrow \lambda r$ ($A\rightarrow \lambda^3 A$),
$T\rightarrow \lambda^{-1}T$ ($\beta \rightarrow \lambda \beta$),
$E \rightarrow \lambda^2 E$, and  $S\rightarrow \lambda^3 S$.
It makes sense that the temperature is no
longer an intensive parameter, due to the fact that
in gravitational systems
thermodynamic equilibrium
is obtained through Tolman's formula for the redshifted temperature.
The same goes for the pressure,
where  with
Eq.~(\ref{euler equation})  one finds that the pressure scales as 
$p\rightarrow \lambda^{-1} p$.
Also, the action $I$ and free energy $F$  scale as
$I\rightarrow \lambda^3 I$ and $F\rightarrow \lambda^2 F$.
Taking the differential of the Euler relation and using the first law,
the Gibbs-Duhem relation is 
\begin{equation}
3 S dT + dE - 3Adp = 0.
\end{equation}

\section{Thermal stability}
\label{secStability}

Thermal stability in the canonical ensemble
is determined by the heat capacity at constant cavity area,
$C_A$, which has to be equal to or greater than zero,
$C_A\geq0$.
$C_A$ is defined by
\begin{equation}
C_A \equiv \left( \frac{\partial E}{\partial T} \right)_A \,.
\end{equation}
$E$ is given by Eq.~(\ref{E})
in the form
$E=E(r_+,r)$, with $r_+=r_+(\beta,r)$
given in Eq.~(\ref{beta}). Now, $A={\rm const}$
means $r={\rm const}$, and since $T=1/\beta$,
see Eq.~(\ref{betat}),
we have
$C_A \equiv \left(
\frac{\partial E}{\partial T}
\right)_A 
 = - \beta^2
\frac{ \left( \partial E / \partial r_+ \right)_r}
{\left( \partial \beta / \partial r_+\right)_r}$.
Then, from Eq.~(\ref{E}) we find that $\left( \frac{\partial E}
{\partial r_+} \right)_r = \frac{3\pi}
{4 \sqrt{1-r_+^2/r^2} } r_+$,
and from 
Eq.~(\ref{beta})
we have $
\left( \frac{\partial \beta}{\partial r_+} \right)_r =
\frac{2\pi}{\sqrt{1-r_+^2/r^2}} \left(1 -
\frac{2 r_+^2}{r^2} \right)$. So, 
the heat capacity is given by
\begin{equation}\label{hc}
C_A = \frac{3\pi^2 r_+^3}{2}
\frac{
\left( 1 -\frac{r_+^2}{r^2} \right)}
{\left( 
\frac{2r_+^2}{r^2} - 1 \right)}.
\end{equation}
Thus. thermal stability,
\begin{equation}\label{cageq0}
C_A\geq0\,,
\end{equation}
means that the cavity's boundary must be located between the horizon
radius and the photon sphere radius, $r_{\rm ph}$,
\begin{equation}\label{castability}
{r_+ \leq r \leq r_{\rm ph}}\,,
\end{equation}
where in five dimensions $r_{\rm ph} = \sqrt{2} \, r_+$
\cite{monteiro}.  The same type of result appears in York's
four-dimensional treatment, i.e., for $r_+ \leq r \leq r_{\rm ph}$ with
$r_{\rm ph}= \frac32 r_+$ in $d=4$, the heat capacity is positive and
the canonical system is thermodynamically stable.  Thus, it is clear
that the the photon sphere plays an intrinsic role in thermodynamic
stability, since it appears consistently as a bound in both four and
five dimensions, see also \cite{GregRoss}.  In addition,
Eqs.~(\ref{hc})-(\ref{castability}) imply that the black hole with
horizon radius $r_{+1}$ given in Eq.~(\ref{bh1}) is always
thermodynamically unstable and the black hole with horizon radius
$r_{+2}$ given in Eq.~(\ref{bh2})  is always thermodynamically
stable.

\section{Free energy function of the five-dimensional black hole:
Thermodynamic 
interpretation of the 
physical processes}
\label{secFreeEnergyFunc}

We can make progress and understand the physical processes and
thermodynamic instabilities further within the thermodynamic approach
using a generalized free energy.

The free energy $F$ in our problem was defined for $r_+=r_{+1}$ and
$r_+=r_{+2}$, i.e., $F=F(r_+,r)$ or $F=F(\beta,r)$ with $r_+$ taking
those two values.  It is of interest to define a generalized canonical
free energy $\bar F$ as ${\bar F}=E-TS$ where the boundary radius $r$
and the temperature $T$ are still kept fixed, but the horizon radius
$r_+$ can assume all values, not just the two above.  Using such an
extension $\bar F$ for the five-dimensional free energy, as was done
in four dimensions in \cite{york1}, one can analyze radial, i.e.,
spherical, and static perturbations to the thermodynamic system we are
studying, thus providing results for the possible stable and unstable
configurations.

\begin{figure}[h]
\centering
\includegraphics[width=0.48\textwidth]{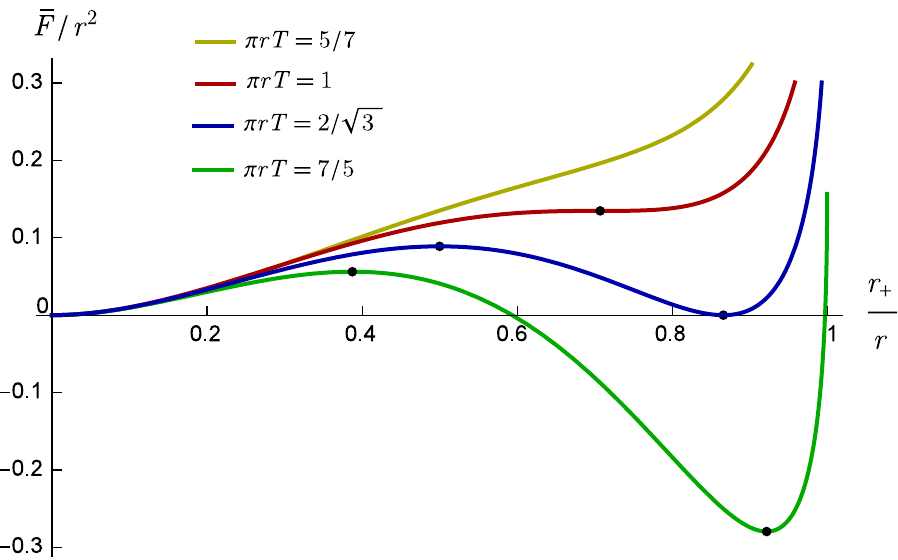}
\caption{The free energy function for four different situations,
namely, $\pi r T=\frac57$, $\pi r T=1$, $\pi rT=\frac{2}{\sqrt{3}}$,
and $\pi rT=\frac{7}{5}$.  Classical hot flat space is at $r_+=0$ and
has zero free energy. The black hole solutions $r_{+1}$ and $r_{+2}$
occur at the extrema $\left( \frac{\partial {\bar F}}{\;\,\partial
r_+} \right)_{r,T} = 0$, indicated on the plot by the black dots.  The
upper yellow curve $\pi rT=\frac57$, an instance of $\pi rT<1$, is for
the case where there are no black hole solutions.  The red curve is
for the limiting case $\pi rT=1$, where the two black hole solutions
coincide at the inflection point, $r_{+1}=r_{+2}$, resulting in a
neutral equilibrium point. As $\pi rT$ increases, the smaller black
hole $r_{+1}$ is an unstable equilibrium point, while the larger black
hole $r_{+2}$ is stable.  The smaller black hole $r_{+1}$ always has
positive free energy, and the larger black hole $r_{+2}$ can have
positive or negative free energy depending on the value of $\pi rT$.
For $\pi rT<\frac{2}{\sqrt{3}}$ the free energy of the larger black
hole $r_{+2}$ is positive, for $\pi rT=\frac{2}{\sqrt{3}}$ the free
energy of the larger black hole $r_{+2}$ is zero, and for $\pi
rT>\frac{2}{\sqrt{3}}$, e.g., $\pi rT=\frac{7}{5}$ the free energy of
the larger black hole $r_{+2}$ is negative, and classical hot flat
space $r_+=0$ which has zero free energy can nucleate in this latter
case into the larger $r_{+2}$ black hole through the $r_{+1}$ black
hole.}
\label{FreeEnergy}
\end{figure}

From Eqs.~(\ref{E}) and~(\ref{finalentropy}) we find that the extended
free energy ${\bar F}=E-TS$ can be written as ${\bar F}={\bar
F}(r_+,r,T)$ and is given by
\begin{equation}
{\bar F}(r_+,r,T) =
\frac{3\pi r^2}{4} \left( 1 - \sqrt{1-\frac{r_+^2}{r^2}} - \frac{2\pi
r T}{3} \frac{r_+^3}{r^3} \right)\,.
\label{barf}
\end{equation}
With
this ${\bar F}$ we are considering static $r_+$ deviations from the
two solutions $r_+=r_{+1}$ and $r_+=r_{+2}$.  In
Fig.~\ref{FreeEnergy} we plot $\bar F$ as a function of $r_+$,
keeping $r$ and $T$ fixed, as given in Eq.~(\ref{barf}).  The
extrema in the figure occur for $\left( \frac{\partial {\bar
F}}{\partial r_+} \right)_{r,T} = 0$. Solving for these
extrema, one finds
that they occur for $r_+=0$, $r_+ = r_{+1}$, and $r_+ = r_{+2}$, i.e.,
for no black hole, and for the two black hole solutions in the
canonical ensemble.  Indeed, the no black hole case $r_+=0$ is a solution
with ${\bar F}=0$. We have been dealing with the classical
gravitational action for the canonical ensemble thermodynamic
approach and the no black hole case corresponds to having a
boundary $r$ at temperature $T$ with nothing inside, only Minkowski,
flat spacetime at temperature $T$.  This is classical hot flat space,
i.e., classical vacuum at temperature $T$ with zero free energy.  From
Fig.~\ref{FreeEnergy} we see that the smaller black hole at $r_{+1}$
is an unstable equilibrium point and it has $C_A <0$, while the larger
black hole $r_{+2}$ is a stable equilibrium point and it has $C_A >0$,
as found previously.  One also finds from $\left( \frac{\partial {\bar
F}}{\partial r_+} \right)_{r,T} = 0$ that there are extrema only when
$\pi r T \geq 1$, with the equality yielding the case $r_{+1}=r_{+2}$, a
result consistent with our findings.
Note that ${\bar F}$ in
Eq.~(\ref{barf}), see also Fig.~\ref{FreeEnergy}, has been defined for
all $r_+$, but for a value of $r_+$ other than $r_+=0$ and the two
black hole values $r_{+1}$ and $r_{+2}$ the geometry has a conical
defect in the real Euclidean plane and the corresponding temperature
does not correspond to a physical situation, i.e., the corresponding
black hole does not have the correct temperature since it does not
have a temperature compatible with the temperature $T$ of the
boundary, and so it is not an equilibrium configuration.

For $\pi r T < 1$ only classical hot flat space, with $r_+=0$, is
possible inside the cavity at radius $r$, as can be deduced from
Eq.~(\ref{barf}) or seen in Fig.~\ref{FreeEnergy}, i.e., it is
impossible to have black holes, see also Eq.~(\ref{real}).  For $\pi r
T \geq 1$ the black hole solutions $r_{+1}$ and $r_{+1}$ are possible,
but to extract the full details a careful analysis with $\bar F$ has
to be performed.  It is useful to resort to Fig.~\ref{FreeEnergy}.
For $\pi r T \geq 1$ one can follow the possibility of forming black
holes from hot flat space through a nucleation instability.  Classical
hot flat space has $r_+=0$ by definition.  The smaller black hole
$r_{+1}$ always has the boundary $r$ obeying $r>r_{\rm ph}$, 
always has positive free energy, i.e., positive action $I(r_{+1})$ given
by Eq.~(\ref{bhactionr+1}), where
from now on we shorten the notation
$I(r_{+1})\equiv I(r,r_{+1}(r,\beta))$,
is at a
maximum of $\bar F$, see Fig.~\ref{FreeEnergy}, and this means
it is unstable, with
the thermal instability also being seen through the negativity of the
specific heat.  The larger black hole $r_{+2}$ always has the boundary
$r$ that obeys $r<r_{\rm ph}$, can have both either
positive or negative free
energy, i.e., positive or negative action $I(r_{+2})$ given by
Eq.~(\ref{bhactionr+2}) with the notation shortened, respectively, is
at a minimum of $\bar F$, see Fig.~\ref{FreeEnergy}, and this means
it is stable, with
the thermal stability also being seen through the positivity of the
specific heat.  The maximum of $\bar F$ at $r_{+1}$ forms a potential
barrier between classical hot flat space at $r_+=0$ and the larger
stable black hole $r_{+2}$ at a minimum of $\bar F$ at $r_{+2}$.  It
is then of interest to analyze for which conditions the stable
solution $r_{+2}$ can be reached through the unstable one $r_{+1}$
from hot flat space $r_+=0$.

It is known that in the canonical ensemble, any spontaneous process
goes in the direction of decreasing $F$. In the canonical ensemble it
is forbidden to increase $F$, in much the same way that it is
forbidden to decrease the entropy $S$, i.e., violate the second law of
thermodynamics in the microcanonical ensemble, where the thermodynamic
energy $E$ and the size of the system are held fixed.  Thus, a phase
transition in the canonical ensemble from classical hot flat space
$r_+=0$ to the larger stable black hole $r_{+2}$ can occur as a
spontaneous process if the $r_{+2}$ black hole has lower free energy than
hot flat space.  Classically, hot flat space has zero free energy, so
if the $r_{+2}$ black hole has negative free energy nucleation can
occur to a black hole $r_{+2}$.  Thus, to have nucleation one requires
$F(r_{+2})\leq0$, i.e., $I(r_{+2})\leq0$, where  the equality
holds when the two phases coexist.  From Eq.~(\ref{bhactionr+2}), this
yields
\begin{equation}
\label{ixx21}
\pi rT\geq\frac{2}{\sqrt{3}}\,,
\end{equation}
which in round numbers is $\pi r T \geq\frac{2}{\sqrt{3}} = 1.1547$.
This is a necessary and sufficient condition for nucleation from
classical hot flat space to the stable black hole $r_{+2}$ through the
unstable black hole $r_{+1}$. Equation~(\ref{ixx21}) is a stronger
condition than the $\pi rT\ge1$ condition of Eq.~(\ref{real}) for
having black hole solutions at all.  Conversely, one finds that a
large black hole $r_{+2}$ decays into classical hot flat space when
the condition $F(r_{+2})\geq0$ holds, i.e., $I(r_{+2})\geq0$.  From
Eq.~(\ref{bhactionr+2}) this yields $\pi rT\leq\frac{2}{\sqrt{3}}$,
which together with the $\pi rT\ge1$ condition for having black holes
gives that a large black hole $r_{+2}$ can decay into classical hot
flat space when \begin{equation}\label{ixx21x} 1\leq\pi
rT\leq\frac{2}{\sqrt{3}}\,.  \end{equation} Also, within this range
classical hot flat space will never nucleate into a black hole.

\section{Action functional to second order and its relation to
thermodynamics and thermodynamic stability}
\label{sa}

In the Hartle-Hawking path-integral approach to quantum gravity
\cite{hawking2,hawking3,hawking4}, the partition function $Z$ for the
gravitational field is given by the integral of the Euclidean Einstein
action $I$ over the space of metrics $\rm g$, i.e., $Z= \int
d\rm [g] \, \exp(-I[\rm g])$.  The path-integral approach of
the canonical ensemble for black holes~\cite{york1} uses the classical
action, i.e., the zero-order Euclidean Einstein action $I_{\rm BH}$,
for the possible black hole solutions together with the corresponding
partition function $ Z= \exp(-I_{\rm BH})$, giving the black hole
thermodynamic properties that we have been analyzing.  We have seen that
for a given temperature $T$ at some cavity radius $r$, there is
classical hot flat space $r_+=0$, and there are two black hole radii,
$r_{+1}$ and $r_{+2}$, these three situations are
the ones that fit the boundary data.
Two important results in this formalism
are a formula for the black hole entropy and
that there can be a phase
transition from classical hot flat space to the larger stable black
hole $r_{+2}$ through the smaller unstable one $r_{+1}$ and vice
versa, as a
detailed analysis of the free energy has shown.

Now, a next-order quantum gravity treatment should
include the fact that hot
flat space, rather than being empty as we have been considered so far,
contains gravitons and so its free energy is nonzero,
in fact is negative. Indeed graviton
radiation in thermal equilibrium in a cavity of radius $r$ at
temperature $T$ is graviton blackbody radiation.
In addition, a Schwarzschild black hole is also made of
gravitons and to understand it in more detail
one should study perturbations
due to quantum thermal fluctuations around the Euclidean
black hole
solutions.

Thus, to proceed with the path-integral formulation and to calculate
the partition function $Z= \int d [\rm g] \, \exp(-I[\rm g])$
anew to include fluctuations around the Euclidean black hole
solutions, also called black hole instantons in this context, one considers
small metric perturbations $h_{ab}$ to the instanton metric ${\bar
g}_{ab}$, such that $g_{ab}={\bar g}_{ab}+h_{ab}$.  The action $I$ is
now $I=I[{\bar g}]+I_2[{h,{\bar g}}]$, where a first-order term does
not appear because the instanton satisfies the equations of motion and
terms with order higher than two are discarded. The partition function
is then $Z={\rm e}^{-I[{\bar g}]} \int d[h] {\rm e}^{-I_2[h,{\bar g}]}=
{\rm e}^{-I[{\bar g}]}Z_2$, for some function $Z_2$.  The perturbed
part $I_2[{h,{\bar g}}]$ reveals the fluctuating modes, their
eigenfunctions, and eigenvalues.  There are two possibilities
depending on the latter.  If in the perturbation one of the modes gets
a negative eigenvalue one has a negative mode around the given
instanton and the perturbation is unstable. This in turn means that
the action and thus the free energy has an imaginary part, and that
the partition function also gets an imaginary exponential. Going back
to the Lorentzian representation, the associated action functional has
a negative exponential signaling a decaying process, i.e., an
instability of the system.  On the other hand, there is the
possibility that all of the perturbed modes have positive eigenvalues,
in this case the
perturbation is stable around the given instanton.

In \cite{gpy}, it was found for a four-dimensional spacetime, that in
the canonical ensemble, i.e., for fixed $T$ at a fixed boundary with
very large radius $r$, perturbations around the instanton $r_{+1}$
yield a negative eigenvalue for which the corresponding mode decreases
the action. So, the instanton is a saddle point rather than a minimum,
and is unstable.  For this small black hole $r_{+1}$ the free energy
then gets an imaginary part which is inherited by the partition
function. Then, the path integral is an exponential with a negative
exponent, yielding a mode with a finite decay time, either to quantum
hot flat space or to the large black hole $r_{+2}$ \cite{gpy}.
In~\cite{allen} it was further shown in four dimensions that, modulo
gauge boundary conditions, the negative mode disappears and the system
is stable when the cavity radius $r$ obeys $r\leq r_{\rm ph}$, where
again $r_{\rm ph}$ is the radius of the circular photon orbits of the
spacetime, i.e., $r_{\rm ph}=\frac32 r_+$ for four-dimensional
Schwarzschild spacetime.  This means that if one calculates the
fluctuations around the instanton $r_{+2}$ one finds that the action
has no negative eigenvalue and the modes increase the action away from
the instanton; thus it has a minimum, the partition function is
real, there are no decaying modes, and the black hole is stable.
Such a perturbation analysis was done 
in~\cite{GregRoss} in
five and higher dimensions for the two Schwarzschild instantons
$r_{+1}$ and $r_{+2}$, and it was indeed found that for a cavity radius
$r$ less than or equal to a given radius the negative mode
disappears. This radius is the radius of the circular photon orbits,
$r_{\rm ph}$, which in five dimensions is $r_{\rm ph} =
\sqrt{2} \, r_+$.

Clearly, one expects the instability provided
by the heat capacity and the free energy thermodynamic phase
transition analysis done previously
to be related to the action functional path
integral instability and the corresponding
phase transitions.  This is
indeed the case.  From the thermodynamic point of view, the small
black hole solution $r_{\rm +1}$ is a maximum of the free energy
function and has negative heat capacity, which means
that it is an unstable
solution.  From the second-order action functional path-integral
analysis, the black hole instanton $r_{+1}$ is a saddle point, as it
has a negative eigenvalue that decreases the action, rendering the
solution unstable.  It acts as a barrier over which hot gravitons can
be excited.  From the thermodynamic point of view, the large black
hole solution $r_{\rm +2}$ is a minimum of the free energy function
and has positive heat capacity, which means
that it is a stable solution.  From
the second-order action functional path-integral analysis, the black
hole instanton $r_{+2}$ is a minimum of the action, it has no negative
eigenvalues, and the solution is stable.  Both points of view show
that hot flat space is metastable in some circumstances
and can transition and nucleate the
large black hole $r_{\rm +2}$ through the barrier provided by the
small black hole $r_{\rm +1}$.  The definite connection between the
thermodynamic heat capacity and the saddle approximation of the
action, i.e., that the specific heat of the black hole becomes
positive when the negative mode of the black hole instanton
disappears, was made clearer in \cite{reallbranes} where a
connection to a classical mechanical instability of black branes
was also made.

\section{Ground state of the canonical ensemble:
Quantum hot flat space,
black hole, or both}
\label{secGround}

The action functional perturbation analysis
has thus showed that
in order to properly understand 
the physics involving black holes
one has to treat hot flat space in
quantum terms rather than in classical
terms. Knowing this, one can address the important
issue of determining  the ground state
of the canonical ensemble.
By definition, the ground state of the canonical
ensemble is the one with the
lowest free energy $F$ or, since $I=\beta F$,
the one with the lowest action $I$.
In the canonical ensemble for hot gravity
that we are studying, 
there are three possible phases: quantum hot flat space,
the large $r_{+2}$ black hole, or a 
superposition of these two phases. 
Thus, to decide
which one is the ground state
we have to compare the free energy of quantum hot flat space
$
F_{\rm HFS}
$
with
the free energy
$F(r_{+2})$
for the stable black hole $r_{+2}$.

First, we determine the
free energy of quantum hot flat space.
In quantum hot flat space
one treats the system
as a quantum gas of gravitons living in a
cavity of radius $r$ at temperature $T$.
Thus, quantum hot flat space is
a blackbody of gravitons at a given temperature.
The internal energy
of free gravitons 
inside a cavity of a four-dimensional
fixed spatial four-volume $V$ of arbitrary shape and fixed
temperature $T$ is 
$
E_{\rm HFS} = 5 a V T^5
$,
where the $5$ in front of the
expression is the total number of
degrees of freedom
of a massless graviton in five dimensions,
and 
$a$ is a number given by 
$a= \frac{3 \zeta(5)}{\pi^2}$, where $\zeta$ is the Riemann zeta 
function with $\zeta(5)=1.03693$
in round numbers, and it is related to the 
Stefan-Boltzmann constant in four spatial dimensions $\sigma$ 
through
$a= \frac{8}{3\pi}\sigma$~\cite{landsberg}.
A spherical cavity of four spatial dimensions has volume 
$V=\frac{\pi^2 r^4}{2}$ and the equation of state for
radiation is
$
E_{\rm HFS} = 4 P_{\rm HFS}\,V
$,
where
$P_{\rm HFS}$ is the radiation pressure
given by $P_{\rm HFS}=\frac54 a  T^5$.
Then, the entropy $S_{\rm HFS}$ follows from the first law
of thermodynamics
applied to the radiation in the cavity,
$TdS_{\rm HFS} = dE_{\rm HFS} + P_{\rm HFS} dV$, i.e.,
$
S_{\rm HFS} = \frac{3 E_{\rm HFS}}{4T}
$.
The free energy of quantum hot flat space can now be calculated from
$F_{\rm HFS}=E_{\rm HFS}-TS_{\rm HFS}$
and it is
\begin{equation}\label{ihfs1}
{F}_{\rm HFS} = - \frac{5a \pi^2}{8} r^4 T^5\,.
\end{equation}
Quantum hot flat space has a free energy
with a $r^4T^5$ dependence,
and it is negative,
rather than zero as in the classical approximation.
In the case where one wants to work with the action
for quantum hot flat space, one has 
$I=\beta F$ with $\beta=1/T$, and so
$
I_{\rm HFS} = - \frac{5\pi^2 a}{8}r^4 T^4
$.

Second, the free energy $F(r_{+2})$
for the 
stable black hole $r_{+2}$
is
given 
by
$I(r_{+2})$ of 
Eq.~(\ref{bhactionr+2}), where again
we have shortened the notation. Since $I=\beta F$,
one gets 
\begin{align}
&F(r_{+2}) = \frac{3 \pi^2 {r}^3T}{4} \Bigg[ (\pi {r}{T})^{-1} 
-\nonumber \\
&-\frac{2\sqrt2}{3}\sqrt{1+\sqrt{1-(\pi {r}{T})^{-2}}} 
\left( 1- \frac{\sqrt{1-(\pi {r}{T})^{-2}}}2 \right) \Bigg].
\label{scaledbhaction}
\end{align}

Now one has to directly compare 
${F}_{\rm HFS}$
given in Eq.~(\ref{ihfs1})
with $F(r_{+2})$ given in Eq.~(\ref{scaledbhaction}).
Comparing both free energies, one has that the stable black hole
is the ground 
state whenever
\begin{equation}\label{iscomp}
F(r_{+2})\leq{F}_{\rm HFS}\,,
\end{equation}
with the equality holding when both phases coexist.

\begin{figure}[t]
\centering
\includegraphics[width=0.48\textwidth]{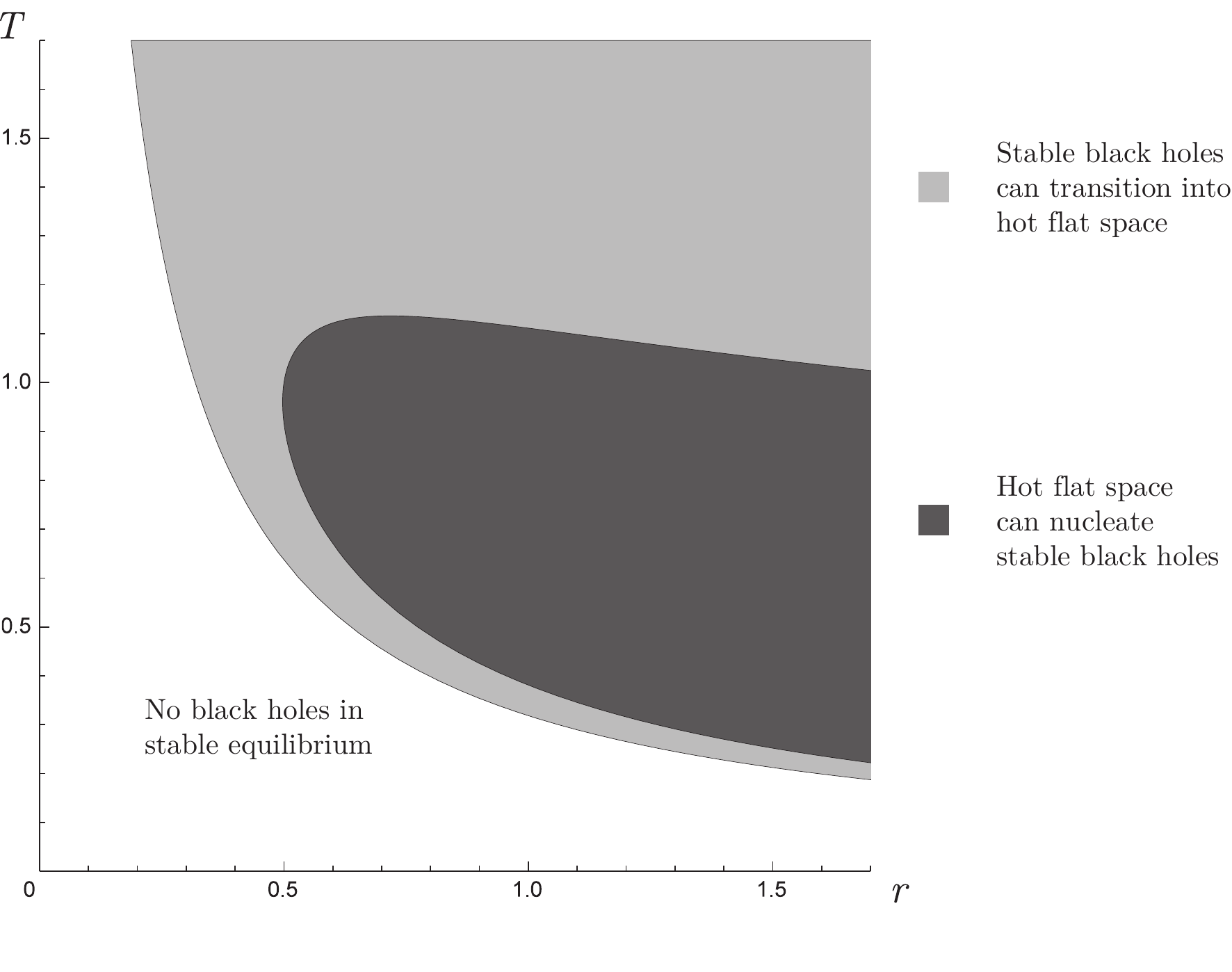}
\caption{
In the $({r},{T})$ plot, each configuration of the canonical ensemble
is represented by a different point.  The white region defined by
$\pi{r}{T}<1$, see Eq.~(\ref{real}), is the region where black holes
are not possible.  The gray region has quantum hot flat space as the
ground state.  The black region has the larger stable black hole
$r_{+2}$ as the ground state of the canonical ensemble.  The thick
black line corresponds to the situation where the ground state is a
superposition of the black hole state and quantum hot flat space.
These regions are separated by the line implicitly defined in
Eq.~(\ref{iscomp}).  It is interesting to comment on the hyperbola
$\pi{r}{T}=\frac{2}{\sqrt3}$ that appears in Eq.~(\ref{ixx21}). This
line is not drawn so as not to confuse the figure. Above this hyperbola the
black hole action is negative and below it the black hole action is
positive. The hyperbola appears naturally when one considers classical
hot flat space with zero free energy rather than quantum hot flat
space with negative free energy.  One finds that for relatively large
$r$, i.e., $r \gtrsim1.0$ this classical line coincides with the thick
black line of the figure, which is the line of coexistence, so
classical hot flat space is a good approximation in this region.  For
relatively small $r\lesssim1.0$ the classical line hyperbola detaches
from the black line of coexistence, meaning the quantum theory for hot
flat space is needed in the region of small $r$ and high $T$.
}
\label{GroundState}
\end{figure}

Using Eqs.~(\ref{ihfs1}) and~(\ref{scaledbhaction}) in
Eq.~(\ref{iscomp}) we find
\begin{align}
&\frac{5\pi^2 a}{8} {r}^{-3} < - \frac{3\pi^3}{4} (\pi {r}{T})^{-4} 
\Bigg[ (\pi {r}{T})^{-1} -\nonumber \\
&-\frac{2\sqrt2}{3}\sqrt{1+\sqrt{1-(\pi {r}{T})^{-2}}} 
\left( 1- \frac{\sqrt{1-(\pi {r}{T})^{-2}}}2 \right)
\Bigg], \label{ground ineq2}
\end{align}
The right-hand side of Eq.~(\ref{ground ineq2})
has a maximum with respect to $\pi
{r} {T}$ at
$\pi {r}{T} = \frac{121}{8\sqrt105}$, i.e.,
$\pi {r} {T} =1.47605$ in round numbers.
For that maximum, again from Eq.~(\ref{ground ineq2}),
there is a corresponding critical radius ${r}_c$
given by 
${r}_c=\frac{1331}{336}
\sqrt[3]{\frac{121\,a}{20\sqrt{105}\pi^4}}$, i.e.,
${r}_c=0.491529$ in round numbers.
If the radius is higher than the critical value ${r}_c$ then the
black hole is the ground state for some cavity temperature
${T}$, if the radius is lower than the critical value
${r}_c$ then the ground state
is quantum hot flat space, and
if there is equality the ground state
is
made of both phases living together, see Fig.~\ref{GroundState}.

We can also use Eq.~(\ref{iscomp}) to find a critical cavity
temperature $T_c$.  Indeed, using again Eqs.~(\ref{ihfs1})
and~(\ref{scaledbhaction}), Eq.~(\ref{iscomp}) can also be written as
\begin{align}
&\frac{5\pi^2 a}{8} {T}^3 \leq -
\frac{3\pi^3}{4} (\pi {r}{T})^{-1} 
\Bigg[ (\pi {r}{T})^{-1} -\nonumber \\
&-\frac{2\sqrt2}{3}\sqrt{1+\sqrt{1-(\pi {r}{T})^{-2}}} 
\left( 1- \frac{\sqrt{1-(\pi {r}{T})^{-2}}}2 \right)
\Bigg]. \label{ground ineq}
\end{align}
The right-hand side of Eq.~(\ref{ground ineq}) has a maximum with
respect to $\pi {r} {T}$ at $\pi {r} {T} = \frac{25}{4\sqrt{6}}$, in
round numbers this is $\pi {r} {T} =2.55155$.  For that maximum, again
from Eq.~(\ref{ground ineq}), there is a corresponding critical
temperature ${T}_c$ given by ${T}_c=\frac{4}{5}
\sqrt[3]{\frac{36\pi}{125a}}$, which with $a=0.315188$ approximately,
gives ${T}_c =1.13697$ in round numbers.  If the temperature is lower
than the critical value ${T}_c$ then the black hole is the ground
state for some cavity radii ${r}$, if the temperature is higher than
the critical value ${T}_c$ then the ground state is quantum hot flat
space made of gravitons, and if the equality holds the ground state is
both phases living together, see Fig.~\ref{GroundState}.

In brief, the canonical ensemble for hot gravity in five dimensions
possesses a ground state that can be either a quantum hot flat space
phase, a large black hole phase, or a superposition of these two
phases.  Which phase dominates is determined by the cavity radius $r$
and temperature $T$, as shown in Fig.~\ref{GroundState}.  If one
changes the cavity radius or temperature, a phase transition from one
phase to the other might occur.  It is worth noting that the phase
transition implies a change of topology of space.  For instance, when
the phase changes from quantum hot flat space to a black hole the
topology changes from $R^5$ of flat space to $R^2\times S^3$ of the
Euclidean black hole.  Now, the Euler characteristic is a measure of
the topology of the space.  In the transition from classical hot flat
space $r_+=0$ to the stable $r_{+2}$ black hole through the $r_{+1}$
black hole, the Euler characteristic $\chi$ of the Euclidean $(t,r)$
disk \cite{btz1994} of the sequence of geometries takes the value
$\chi=0$ for classical flat space at $r_+=0$ and the value $\chi=1$ at
$r_{+1}$ and $r_{+2}$.  The Euler characteristic is not defined for
other values of $r_+$.  Therefore, in the transition there is a change
of spatial topology, a fact that is known to be possible in quantum
gravity at finite temperature but impossible in classical gravity.
It is also interesting to note that we have found that in the Planckian
regime, namely, $r=1$ and $T=1$ the black hole is the ground state of
the ensemble.  This result is altered if quantum hot flat space has
many additional fundamental species on top of the five graviton
degrees of freedom; in such a case, the ground state at the Planck
scale is quantum hot flat space.

\section{Density of states}
\label{secDensity}

We now show that by using the partition function representation as a
sum over energies $E$ of the density of states $\nu$ weighted with the
Boltzmann factor $e^{-\beta E}$, one can verify that $\nu$ is
proportional to the logarithm of $\frac{A_{+2}}{4}$, establishing that
the entropy of the $r_{+2}$ black hole is indeed $S=\frac{A_{+2}}{4}$.

In the canonical ensemble, the
partition function $Z(\beta,r)$
can be written as an integral
over the energies of the system in the form
$Z(\beta,r)=\int dE \,\nu(E,r)e^{-\beta E}$,
with $\nu(E,r)$ being the density of states
at given $E$ and cavity radius $r$.
Conversely, given a partition function $Z(\beta,r)$
one can calculate for a given $r$
the number of states $\nu(E,r)dE$ 
between $E$ and $E+dE$ using the Laplace transform
$
\nu(E,r) = \frac{1}{2\pi i} \int_{-i \infty}^{i \infty}
d\beta Z(\beta,r)e^{\beta E}
$.
For the stable black hole $r_{+2}$, the partition function reads
$Z=e^{-I(r_{+2})}$,
where $I(r_{+2})$ is given in Eq.~(\ref{bhactionr+2}).
Since for the existence of black holes one has $\pi r T
>1$, see  Eq.~(\ref{real}),
we can  approximate $r_{+2}$, see  Eq.~(\ref{bh2}),
to second order in $1/(\pi r T)^2$
giving
$r_{+2} = r \left( 1 - \frac{1}{8 \left(\pi r T \right)^2} +
\mathcal{O}(T^{-4}) \right)$, and then
with 
$I(r_{+2})$, see Eq.~(\ref{bhactionr+2}),
to second order in $1/(\pi r T)^2$
we can write the partition function~(\ref{zeroloop})
as
\begin{equation}
Z(\beta,r) \simeq \exp \left( \frac{\pi^2}{2}r^3 -
\frac{3\pi}{4}\beta r^2 + 
\frac{3}{16}\beta^2 r  \right).
\end{equation}
Putting this into
the Laplace transform
$
\nu(E,r) = \frac{1}{2\pi i} \int_{-i \infty}^{i \infty}
d\beta Z(\beta,r)e^{\beta E}
$
one finds that the density of states
is 
$\nu(E) = \frac{2}{\sqrt{3\pi r}}
\exp \left(\frac{\pi^2 r^3}{2} - \frac{4}{3 r} 
\left(E-\frac{3\pi r^2}{4} \right)^2 \right)$.
Using from
Eq.~(\ref{E}) that the ADM mass $m$ is
$m = E - \frac{2 E^2}{3 \pi r^2}$, 
one is able to put $\nu(E)$ in terms of $m$, i.e.,
$
\nu(E) = \frac{2}{\sqrt{3\pi r}}
\exp \left(- \frac{\pi^2 r^3}{4} + 2 \pi r m \right)
$.
Also, since in this approximation the cavity radius and the event
horizon are very close to each other, one can take $r = r_{+2}$, 
use $m=\frac{3\pi}{8}r_+^2$, and find that the density of states is
well approximated by
\begin{equation}\label{ds1}
\nu(E) = \frac{2}{\sqrt{3\pi r}}\,\,
{\rm e}^{\frac{A_{+2}}{4}},
\end{equation}
where $A_{+2}= 2\pi^2 r_{+2}^3$ is the black hole's
surface area.
In statistical mechanics 
the relation between  entropy $S$ and the
density of states $\nu$ is
$S=a \ln \nu$, for some factor $a$, so we find
that
the black hole entropy is
\begin{equation}\label{entagain}
S=\frac{A_{+2}}{4},
\end{equation}
as one would expect.
Interestingly, this result cannot be obtained if instead of 
$r_{+2}$ one employs the instanton $r_{+1}$
and uses $I(r_{+1})$ as given in Eq.~(\ref{bhactionr+1})
in the partition function 
$Z=e^{-I(r_{+1})}$
and then calculates the Laplace transform for $\nu(E,r)$,
since the integral is divergent. 
Only the large black hole 
$r_{+2}$ makes sense in this setting.

\section*{Acknowledgments}
We thank Charles Robson for conversations.
RA acknowledges support from the Doctoral Programme in the Physics
and Mathematics of Information (DP-PMI) and the Funda\c c\~ao para a
Ci\^encia e Tecnologia (FCT Portugal) through Grant
No.~PD/BD/135011/2017.  JPSL acknowledges FCT for financial
support through Project~No.~UIDB/00099/2020.

\vskip 1cm
\appendix

\section{Connection to York's four-dimensional results}
\label{secappa}

In extending York's four-dimensional canonical black hole
thermodynamics results to five dimensions, there are several points
that are worth mentioning.

When working out the possible black hole radii for a given cavity
radius $r$ and a given cavity temperature $T$, one finds that in four
dimensions one needs to solve
a cubic polynomial with two real radii,
yielding the two black holes.  On the other hand, in five dimensions
one instead
has a quartic potential, and one would expect a different
number of real solutions, but there are also only two, with radii
$r_{+1}$ and $r_{+2}$, yielding again two possible black holes.

The formalism also shows that the two black holes, the small $r_{+1}$
and the large $r_{+2}$, both have a Bekenstein-Hawking entropy in
four and five dimensions, although the whole procedure is only well
defined for the larger one since it is stable and the laws of
thermodynamics can be applied, whereas the smaller one is unstable and
cannot be treated through thermodynamics.

There are two distinct characteristic radii that appear naturally in
the canonical black hole thermodynamics. One is the photon sphere
radius $r_{\rm ph}$.  In four dimensions, for a given temperature $T$
of the heat bath, the photon sphere separates systems that are
thermodynamically stable from systems that are unstable. In five
dimensions the photon sphere also plays this decisive role, namely,
the black holes are in stable equilibrium with the surrounding heat
bath if their cavity radius lies inside the photon sphere radius.
Being a characteristic that seems to appear in any dimension, this shows
that there is some intrinsic property of the photon sphere that
connects it to thermodynamic stability. However, a full explanation
has not been provided yet.  The other characteristic radius that
appears naturally in these thermodynamic systems is the Buchdahl
radius $r_{\rm Buch}$. The Buchdahl radius is the minimum radius a
spherically symmetric interior solution with Schwarzschild exterior can
have, under certain general conditions.
Remarkably, we have shown that
this radius also appears as the radius for which the free energy of
the stable black hole passes through zero. This happens in
both four
and five dimensions. Since the two junctures at which $r_{\rm Buch}$
appears are completely different, it seems that $r_{\rm Buch}$ signals
as well some intrinsic property of the spacetime geometry.

\end{document}